\documentclass{article}
\usepackage[utf8]{inputenc}
\usepackage{amsmath}
\usepackage{amsfonts}
\usepackage{amssymb}
\usepackage{graphicx}
\usepackage{booktabs}
\usepackage[capposition=top]{floatrow}

\usepackage{setspace}
\usepackage{geometry}
 \usepackage{sectsty}
\allsectionsfont{\normalsize}

\usepackage[document]{ragged2e}
\usepackage[numbers]{natbib} 

\begin{document}

\author{Elizabeth A. Handorf*\textsuperscript{1}, Marc Smaldone\textsuperscript{2}, Sujana Movva\textsuperscript{3}, Nandita Mitra\textsuperscript{4}}
\title{Analysis of survival data with non-proportional hazards: A comparison of propensity score weighted methods}
\maketitle

[1]Biostatistics and Bioinformatics Facility, Fox Chase Cancer Center, PA USA

[2]Department of Surgical Oncology, Fox Chase Cancer Center, PA, USA

[3]Department of Medicine, Memorial Sloan Kettering Cancer Center, NY, USA

[4]Division of Biostatistics, University of Pennsylvania Perelman School of Medicine, PA, USA

*Correspondence to: 
Elizabeth Handorf. 333 Cottman Ave., Reimann 383, Philadelphia, PA 19111. 
elizabeth.handorf@fccc.edu

\maketitle

\begin{abstract}
   One of the most common ways researchers compare survival outcomes across treatments when confounding is present is using Cox regression.  This model is limited by its underlying assumption of proportional hazards; in some cases, substantial violations may occur.  Here we present and compare approaches which attempt to address this issue, including Cox models with time-varying hazard ratios; parametric accelerated failure time models; Kaplan-Meier curves; and pseudo-observations. To adjust for differences between treatment groups, we use Inverse Probability of Treatment Weighting based on the propensity score. We examine clinically meaningful outcome measures that can be computed and directly compared across each method, namely, survival probability at time $T$, median survival, and restricted mean survival. We conduct simulation studies under a range of scenarios, and determine the biases, coverages, and standard errors of the Average Treatment Effects for each method. We then apply these approaches to two published observational studies of survival after cancer treatment. The first examines chemotherapy in sarcoma, where survival is very similar initially, but after two years the chemotherapy group shows a benefit.  The other study is a comparison of surgical techniques for kidney cancer, where survival differences are attenuated over time.
\end{abstract}

\section{Introduction}

Many health care studies use observational databases to compare censored survival times between different treatment or exposure groups.   A common statistical approach used to model such data is the Cox proportional hazards regression model. However, the standard Cox model makes a strong assumption that the hazards are proportional between treatment groups of interest over the course of follow-up; that is, that the ratio of the hazards is constant over time. \cite{klein_survival_2003}  This assumption may not hold in real-world studies.
  
%*** lit review of nonPH
 Although Cox regression is often used to analyze survival time data, particularly when adjusting for covariates, there are a variety of models for censored outcomes which do not rely on the assumption of proportional hazards. Non-parametric methods, including the ubiquitous Kaplan-Meier method, by definition do not make assumptions about the functional form of the survival curves, and therefore accommodate non-proportionality. \cite{kaplan_nonparametric_1958}  An alternative non-parametric method is the Nelson-Aalen estimator, \cite{collett_modelling_2015} and a recent method proposes using pseudo-observations, which can also be used without parametric assumptions. \cite{andersen_regression_2004,andersen_pseudo-observations_2010, andersen_causal_2017} However, standard implementations of non-parametric approaches do not readily incorporate covariates.  This motivates use of regression-based parametric and semi-parametric approaches which accommodate non-proportionality.  The Cox model can be modified to include time dependent hazards, relaxing the assumption of proportionality. \cite{collett_modelling_2015}  Accelerated Failure Time (AFT) models are a different class of regression methods for censored data, which assume a linear model for log-time. They are often modeled fully parametrically, and for many choices of distribution for the error term, the models are non-proportional. Semi-parametric AFT methods are also available. \cite{jin_least-squares_2006}  Proportional odds models can also be used in the setting of non-proportionality. \cite{collett_modelling_2015} 

Given the number of choices, a natural question is which method performs best in real-world conditions, particularly when analyzing observational data where  confounding is an issue.  In this paper, we focus on approaches which can use propensity score weighting to address confounding. The propensity score is commonly used in the analysis of clinical studies and has many advantages as we discuss in detail in section 2. We then describe several methods for analysis of survival data where the proportional hazards assumption is violated.  We focus on meaningful estimands and compare the performance of parametric, semi-parametric, and nonparametric approaches; some of which are commonly used while others are less familiar. We focus on methods which can 1) be used with Inverse Probability of Treatment Weighting (IPTW), 2) identify the estimands of interest, %($\Delta S_t$, $\Delta T_{50}$, and $\Delta \mathrm{RMS}_t$)%  
and 3) are readily implemented using currently available software.   We provide insight on the performance of these methods in scenarios often involved in clinical studies. 

This work is motivated by two of our recently published studies in cancer in which the treatment effect exhibited non-proportional hazards. In our  study of chemotherapy for soft-tissue sarcoma, there was a negligible survival benefit until about two years after the start of treatment. \cite{movva_patterns_2015} In another study comparing two different surgical techniques in early-stage renal cancer, we found that one approach had a strong benefit immediately after surgery, but the benefit decreased over time. \cite{ristau_partial_2018}  The data for both of these studies came from an observational source, the National Cancer Database. \cite{boffa_using_2017}  Therefore, both were likely subject to confounding by indication; that is, the choice of treatment is informed by factors that are also associated with survival. 

The outline of our paper is as follows. In section 2, we provide a brief review of estimating the average treatment effect using propensity score weighting. In section 3, we present several methods to analyze survival outcomes under non proportional hazards that utilize IPTW. In section 4, we conduct extensive simulation studies to compare the methods presented in section 3. 
In section 5, we use our motivating studies of soft-tissue sarcoma and renal cancer to demonstrate implementation and interpretation of the various methods described earlier. We conclude with a brief discussion and thoughts on next steps. 
\section{Propensity Score weighting}

The propensity score, first proposed by Rosenbaum and Rubin, \cite{rosenbaum_central_1983, rosenbaum_bias_1984} is defined as the probability of receiving treatment given a set of covariates,
  \begin{equation*}
 e = P(Z=1 \mid \mathbf{X})
 \end{equation*}   
where $e$ is the propensity score, $Z$ is a binary indicator for treatment and $\mathbf{X}$ is a vector of (measured) covariates. The propensity score is useful for causal inference, under the potential outcomes framework.  Here, each subject has two potential outcomes under control and treatment conditions $(Y^0,Y^1)$, one of which is unobserved. \cite{rubin_causal_2005, holland_statistics_1986} The estimand of interest is often the Average Treatment Effect (ATE), defined as
  \begin{equation*}
\mathrm{ATE} = \mathrm{E}(Y^1-Y^0),
 \end{equation*}   
which is usefully interpreted as the expected value of the outcome if every subject is given the treatment, compared with the expected value if every subject receives the control. Strongly ignorable treatment assignment occurs under the conditions 
%*** subscript with i? 
\begin{equation*}
(Y^0,Y^1) \perp Z \mid \mathbf{X} \;
\mathrm{and} \;
0< e < 1  ,
 \end{equation*}   
that is, when the potential outcomes are independent of treatment assignment conditional on the covariates, and each subject has a nonzero probability of receiving both the treatment and the control (positivity). When strong ignorability holds, the ATE based on the propensity score is identifiable from the observed data. \cite{rosenbaum_central_1983, rosenbaum_association_1984, rosenbaum_observational_2002}

\begin{equation*}
\mathrm{E}(Y^1-Y^0) = \mathrm{E}(Y \mid Z=1, e) - \mathrm{E}(Y \mid Z=0, e).
 \end{equation*}   

Propensity score methods have a number of benefits over traditional regression adjustment. First, they provide an easy way to check for sufficient covariate overlap and balance between the treated and control groups. \cite{austin_introduction_2011} Second, one can use flexible and modern approaches, such as ensemble models, \cite{van_der_laan_mark_j_super_2007} to estimate the propensity scores, which make the estimates less vulnerable to bias from an incorrect functional form.  Propensity scores also allow for the use of doubly robust methods. \cite{funk_doubly_2011} Third, propensity score weighting approaches provide marginal, not conditional, estimates, which are more interpretable and useful to clinicians and policymakers. \cite{austin_moving_2015, lunceford_stratification_2004, austin_performance_2017, mao_propensity_2018} 

There are several ways to use propensity scores in analyses of survival data.  Here we focus on weighting, specifically inverse probability of treatment weighting (IPTW).  The idea underlying IPTW is to create synthetic samples where covariates are unrelated to treatment assignment.  The weighted samples can then be summarized and compared directly.  One nice consequence of using IPTW is that the estimand is the ATE. \cite{austin_introduction_2011} To implement an IPTW analysis, one must first obtain the estimated propensity scores for each subject ($\hat{e}_i$).  This can be accomplished via a number of approaches, including logistic regression, random forests, or ensemble methods. The IPTW for subject $i$ is then defined as

\begin{equation*}
\hat{w}_i  = \frac{Z_i}{\hat{e}_i } + \frac{1-Z_i}{1-\hat{e}_i}.
\end{equation*}     
As $Z_i$ is binary, if subject $i$ is treated, their weight is one over their probability of being treated (based on their covariates).  If the subject did not receive the treatment, their weight is one over their probability of receiving the control.  Therefore treated subjects who were less likely to be treated have larger weights, as do control subjects who were less likely to receive the control, thus creating a balanced pseudo-population. In subsequent analyses, these weights are used to estimate the expected effects under the treatment and control conditions.

\section{Survival methods under non proportional hazards}

\subsection{Quantifying differences between groups with non-proportionality} 

%*** aftgee - semi-parametric AFT.  I remember considering this.  I think that it doesn't provide an easy way of getting S(t) 

In the setting of survival outcomes with non-proportionality, we must first decide what our estimand will be.  It should be readily interpretable by the intended audience, meaningful given the observed data features, and estimable using available software. If we fit a proportional hazards model to non-proportional data, we would obtain an estimate of a single HR, but this would be a misleading measure because it would essentially be a summary of the different hazard ratios at each observed failure time.  Alternative hazard ratios have been proposed including the average hazard ratio which is interpreted as the average of the hazard ratios weighted by the number of patients at risk. \cite{schemper_estimation_2009} 

The difficulties inherent in reporting hazard ratios in the setting of non-proportionality can be avoided by focusing on measures that can be defined directly from the survival functions for the treated and control arms, $S_1(t)$ and $S_0(t)$, respectively.

The simplest approach compares differences in survival probabilities at a particular time $t=T$. \begin{equation*}
    \Delta S_t = S_1(t=T) - S_0(t=T)
\end{equation*} 
Quantiles of the survival functions can also be compared. For instance, if the survival curves are defined for $S(t)=0.5$, median survival times can be compared.  
\begin{equation*}
    \Delta T_{50} = (t : S_1(t=T)=0.5) - (t : S_0(t=T)=0.5)
\end{equation*}
These measures have the benefit of being readily understood by a clinical audience, but the drawback is that they do not comprehensively measure survival effects over the length of follow-up.  An alternative measure is  restricted mean survival (RMS), the mean survival up through some time $T$ where $S(t)$ is defined (i.e. mean survival up through the maximum observation time). \cite{conner_adjusted_2019} RMS can be found using the area under the survival curve, where the difference in RMS between the treated and control conditions is
\begin{equation*}
    \Delta \mathrm{RMS}_t  = \int_0^t S_1(T=t)dt - \int_0^t S_0(T=t)dt.
\end{equation*}

Each of these measures are estimated directly from the survival curve, so they are always meaningful regardless of proportionality. However, not every method for the analysis of survival data explicitly estimates the survival curve, and not all methods can accommodate weighting.  In the subsequent sections, we focus on methods which can 1) accommodate IPTW, 2) allow the estimation of the measure of interest ($\Delta S_T$, $\Delta T_{50}$, and $\Delta \mathrm{RMS}_T$), and 3) are readily implemented using published software.

\subsection{Cox model}

Although the Cox proportional hazards model does rely on the assumption of proportionality, the model tends to be robust to small deviations from this assumption. However, if $n$ is large, as is common is registry studies, small deviations from proportionality may be statistically significant when tested using standard diagnostic tests (e.g. Cox-Snell residuals, Schoenfeld residuals). \cite{grambsch_proportional_1994} Therefore, the Cox model may still be useful in some cases were deviations from proportionality are demonstrated. The Cox model is defined as follows:
  \begin{equation*}
    h_i(t \mid Z_i ) = h_0(t)\mathrm{exp}( \beta Z_i ).
  \end{equation*}
IPTW can be readily accommodated in the Cox model, by weighting the contribution of each observation to the partial likelihood function:

  \begin{equation*}
    L(\beta) = \prod_{i=1}^{N} \left( \frac{\mathrm{exp}(\beta Z_i )}{\sum_{j \in R(t_i)} \hat{w}_j \mathrm{exp}(\beta Z_j ) } \right)^{\hat{w}_i}
  \end{equation*} 
where $R(t_i)$ is the risk set at time $t_i$.  The parameter $\beta$ is the average of the log-hazard ratios at each failure time.  Importantly, one can find $\hat{S}(t)$ based on the estimates $\hat{h}_0(t)$ and $\hat{\beta}$.

We can also extend the standard Cox model, relaxing the assumption of proportionality, by allowing the hazard ratio to vary as a function of time.

  \begin{equation*}
    h_i(t \mid Z_i ) = h_0(t)\mathrm{exp}( \beta Z_i +f(Z_i,t)).
  \end{equation*}
In our simulations, we consider two particular cases, 1) allowing the hazard to vary by log-time.

  \begin{equation*}
    f(Z_i,t)=\kappa Z_i \mathrm{log}(t),
  \end{equation*}
and 2) using a piecewise constant treatment effect  
\[  f(Z_i,t)= \left\{
\begin{array}{ll}
0 & 0\leq t < C_1 \\
      \kappa_1  Z_i & C_1\leq t< C_2 \\
      \kappa_2  Z_i & C_2\leq t \\
\end{array} 
\right. \]
These models can all be implemented with the widely-used \texttt{survival} package in R. \cite{terry_m_therneau_package_2020}

\subsection{Parametric AFT models}

Accelerated Failure Time (AFT) models assume a linear model for $\mathrm{log}(T)$. \cite{collett_modelling_2015, jin_least-squares_2006, wei_accelerated_1992}
\begin{equation*}
\mathrm{log}(T_i) = \mu + \beta  Z_i + \sigma \epsilon_i
\end{equation*}
Here, we consider parametric models, where the error term $\epsilon$ is assumed to take on a given distribution.  We have many choices for the distribution of $\epsilon$.  A common choice is the Gumbel distribution, which yields a Weibull hazard model; however, the Weibull model is also a proportional hazards model. We can allow for non-proportionality by allowing the hazard ratio to change over time 

\begin{equation}
    h_i(t \mid Z) =\mathrm{exp}[-(\beta  Z_i +f(Z_i,t))]^\gamma
  \lambda\gamma t^{\gamma -1}  ,
  \end{equation}
where $\lambda $ is the scale and $\gamma$ is the shape parameter. If we let
\begin{equation*}
 f(Z_i,t)=\kappa  Z_i \mathrm{log}(t),
\end{equation*}
then (1) can be re-written as
\begin{equation*}
\mathrm{log}(T_i) = \frac{1}{\gamma + \kappa Z_i}  \left( -\mathrm{log}(\lambda) + \beta Z_i + \epsilon_i \right) .
\end{equation*}
\begin{equation*}
    \epsilon_i \sim  \mathrm{Gumbel} 
\end{equation*}
Therefore, if we include an additional parameter in the Weibull model, where the shape can vary by treatment, the model specification allows the hazard ratio to vary by log-time.

Another flexible alternative is the generalized gamma model, which is also a three parameter model. \cite{marshall_new_1997} The distribution of the errors has an additional parameter $Q$.
\begin{equation*}
\mathrm{log}(T_i) = \mu + \beta  Z_i + \sigma \epsilon_i
\end{equation*}
\begin{equation*}
\epsilon_i \sim  \left( \frac{|Q|}{\Gamma(1/Q^2} \right)
\left( \frac{\mathrm{exp}(Qv)}{Q^2}  \right)^{1/Q^2} 
\mathrm{exp} \left( \frac{-\mathrm{exp}(Qv)}{Q^2} \right)
\end{equation*}

This model encompasses other distributions as special cases; when $Q=1$ we have a Weibull AFT model, and if $Q=0$ it results in a log-normal model. 

For any AFT model, the weights $\hat{w}_i$ can be applied during estimation of parameters. Here, we focus on the 3-parameter Weibull and generalized gamma AFT models, but a wide range of AFT models can be fit in R using the \texttt{flexsurv} package. \cite{jackson_christopher_flexsurv_2016}

\subsection{Kaplan-Meier method}
Another strategy is to avoid parametric assumptions entirely, and use Kaplan-Meier curves with IPTW to directly estimate the survival functions. \cite{cole_adjusted_2004, xie_adjusted_2005} 
\begin{equation*}
\hat{S}(t) = \prod_{t_j \leq t} \left( 1-\frac{\widehat{d_j^w}}{\widehat{n_j^w}} \right)
\end{equation*}
Where $\widehat{d_j^w} = \sum_{i: T_i = t_j}  \hat{w}_i \delta_i$ is the weighted number of events and $\widehat{n_j^w}  \sum_{i: T_i \geq t_j}  \hat{w}_i $ is the weighted number of people at risk at time $t_j$. Weighted curves are easily estimated in R via the \texttt{survival} package.

\subsection{Pseudo-observations}

Another non-parametric method which can be used to estimate the survival curve is the recently-developed pseudo-observations approach. \cite{andersen_regression_2004, andersen_pseudo-observations_2010, andersen_causal_2017} This framework uses a missing data approach to account for censoring.  If no missing data or confounders are present, the ATE can be found using a simple average over the (possibly transformed) outcomes for each subject
\begin{equation*}
\theta = E(f(Y)) = \frac{1}{n}\sum_i f(Y_i).
\end{equation*}
But if $f(Y_i)$ is unknown for some subjects, we cannot calculate $E(f(y))$ directly. Instead, we can define the pseudo-observation for subject $i$ as
\begin{equation*}
  \hat{\theta_i} = n \hat{\theta} - (n-1) \hat{\theta}^{-i}
\end{equation*}
where $\hat{\theta}$ is a consistent estimate for $\theta$, and $\hat{\theta}^{-i}$ is the estimator applied to all observations excluding subject $i$. The pseudo-observations are then used in place of all observations, not only the ones which are missing due to censoring.

In a survival context, our parameter of interest, $\theta$, is the probability of survival at time $t$, and $\theta_i$ is a binary indicator for the status (alive/dead) of subject $i$ at time  $t$. We can use a non-parametric method, such as Kaplan-Meier, to estimate $S_Z(t)$, which is then used to find $\theta_i$ for all observations in arm $Z$.
  \begin{equation*}
  \hat{\theta}_{Zi} = n_Z \hat{S}_Z(t) - (n_Z-1) \hat{S}_Z(t)^{-i}
\end{equation*}
Heuristically, $n_Z \hat{S}(t)$ is the expected number of patients alive at time $t$, and $ (n_Z-1) \hat{S}(t)^{-i}$ is the expected number of patients other than patient $i$ alive at time $t$. $\hat{\theta_i}$ is used in place of (possibly unobserved) survival status at time $t$. Therefore, the probability of being alive at time $t$ in arm $Z$ is
\begin{equation*}
  \hat{\mathrm{P}}(Y|Z, t) = \frac{1}{n_Z}\sum_i \hat{\theta}_{Zi}. 
\end{equation*}

Incorporating propensity-score weights to account for confounding is straightforward. 
\begin{equation*}
  \hat{\mathrm{P}}(Y|Z, t) = \frac{1}{n_Z}\sum_i \hat{w}_i \hat{\theta}_{Zi} 
 \end{equation*}
There are several benefits of using the pseudo-observations approach. It has been demonstrated to provide a valid estimate of the ATE in a causal framework, and requires weaker assumptions than regression-based models. It also supports estimation of many outcomes of interest. We can estimate $\hat{\mathrm{P}}(Y|Z, t)$ at each failure time, defining the survival curve, which in turn allows us to estimate the measures of interest described in section 3.1. Pseudo-observations can also be used to estimate RMS. %We can also use repeated measures logistic regression to estimate hazard ratios. 
This method can be implemented using the R package \texttt{pseudo}. \cite{pohar_perme_maha_pseudo_2017}

\subsection{Variance estimation}
To conduct inference for the outcomes of interest, we need to quantify uncertainty.  Some methods have closed-form variance estimators for our estimands of interest, but others do not. This motivates the use of a non-parametric bootstrap to estimate variances and find confidence intervals.\cite{efron_introduction_1993}  Bootstrap-based estimates are available for every combination of model and outcome we discuss above.  Further, to correctly account for the variance of $\hat{e}$, the estimated propensity scores, we can re-estimate the propensity score within each bootstrap iteration. \cite{austin_variance_2016}

\section{Simulation Studies}

\subsection{Simulation Methods}

We tested the performance of the methods described above using simulation studies.  We assessed the bias, coverage, and variance of each method's estimates of our outcomes of interest: median survival, restricted mean survival, and survival probability at time $T$. We based our covariate and outcome distributions, and effect sizes on our NCDB cancer studies to mimic real world scenarios.

First, we simulated a set of covariates based on the observed multivariate distribution of the NCDB covariates in the renal dataset.  We used the R package \texttt{genOrd} \cite{barbiero_alessandro_genord_2015} to draw variables representing gender, age, stage, histology, tumor grade, Charlson comorbidity score, race, ethnicity, insurance status, facility type, income, and education. As our goal here is to evaluate the survival analysis methods, and not the propensity score estimation methods, we held this set of covariates constant for each simulation; we employ simulated covariates instead of the actual covariates so that our data and simulation code can be shared. \\ (Available online at \verb+https://github.com/BethHandorf/NonPH_IPTW+)

We used a simple logistic model with linear covariate effects to determine each subjects' probability of being assigned to the treatment condition (versus control).  The effect size (log-scale) for each covariate ranged from 0.8 (effect of high vs low stage) to 0.03 (for each additional year of age). (See Appendix Table 1 for full specification.) This corresponds roughly to the effect sizes estimated when modeling actual treatment allocation in the NCDB sample. This model defined the probability that each individual would be allocated to the treatment condition.  We set the intercept such that, on average, the population's probability of receiving the treatment was 0.5.

Survival times were drawn based on a Weibull model with a time-varying treatment effect.  
\begin{equation}
h(t)  = \gamma \lambda t^{\gamma -1} \mathrm{exp}(\beta_1 Z + f(Z, t) + \mathbf{X \psi })
\end{equation}
In our base case, the hazard ratio varied by log time $f(Z, t) = \beta_2 Z \mathrm{log(t)}$, we set $\beta_1$ = -0.69, $\beta_2$ = 0.25, with a sample size of 5000 total cases. For each individual, survival times under $Z=1$ and $Z=0$ were both drawn using the R package \texttt{simsurv}. \cite{sam_brilleman_simsurv_2019} These true potential outcomes were used to calculate the true ATE for the estimands of interest. Observed treatment status was drawn from the binomial based on each individuals probability of receiving the treatment, as defined by our propensity score model. Covariate effect sizes ranged from 0 to 0.8 (see Appendix Table 1).

For each set of simulated outcomes, we first estimated the propensity scores using a logistic regression model, and then calculated respective IPTWs for each individual.  Using these estimated weights, we fit each of the models described above in Section 3 using standard R packages, including \texttt{flexsurv} for parametric survival models and \texttt{pseudo} for pseudo-observations methods.

We repeated this simulation framework with various changes to Equation 2, for a total of five scenarios:

\begin{enumerate}
  \item Base case:  $\beta_1$ = -0.69 ; $f(Z, t) = 0.25 Z \mathrm{log}(t)$
  \item Piecewise constant (PWC) hazard ratio:  $\beta_1$ = 0; $f(Z, t) = -0.25 Z \mathrm{I}(t \geq 2)$, where $\mathrm{I}(\cdot)$ is the indicator function
  \item Modest non-proportionality $\beta_1$ = -0.69 ; $f(Z, t) = 0.125 Z \mathrm{log}(t)$
  \item Modest treatment effect $\beta_1$ = -0.41 ; $f(Z, t) = 0.25 Z \mathrm{log}(t)$
  \item Base case, with a smaller sample size (N=500 per arm)
\end{enumerate}

Senario 1 (base case) had a large treatment effect with substantial non-proportionality. Scenarios 2-4 were chosen to approximate the parameters and non-proportionality found in the clinical examples, and scenario 5 was used to assess how the methods perform with a more modest sample size.

\subsection{Simulation Results}

For each method, we assessed performance based on the mean bias, standard error, and coverage for the  ATE for treatment versus the control. For the base case, the bias was often largest for the standard Cox model, as expected given the substantial degree of non-proportionality in the simulated data. For example, considering median survival, bias for the Cox model was -0.347 years, while all other methods had biases less than -0.069. The generalized gamma AFT model also had large bias relative to other methods; at 10 years, it's bias was actually greater than that of the Cox model (2.5\% vs 2.0\%, respectively).  Even though this three-parameter model is flexible and allows for non-proportionality, fitting this mis-specified parametric model still introduced substantial biases into the results, often comparable in size to those found using the standard Cox model. In the base case, the Cox model with a treatment effect varying by log-time was correctly specified, as was the fully parametric AFT model with variable location and scale parameters.  These models had lower biases than the standard Cox model (e.g. for median survival, biases were -0.042 and -0.056, respectively).  However, the non-parametric Kaplan-Meier and pseudo-observation methods generally had the lowest biases (e.g. biases for median survival were -0.021 and -0.020, respectively).  Finally, we note that the RMS was less sensitive to the choice of method than the other outcomes of interest; the relative differences in the sizes of the biases were smallest for this survival outcome. (see Figure 1)

As shown in Table 1, coverage was worst for the standard Cox model, followed by the generalized gamma AFT model. The time-varying Cox models did better, although coverage was surprisingly low for the log-time model at the 2-year outcome (0.82).  Coverage for the non-parametric methods was close to the nominal level for all outcomes.  Standard Errors (SEs) of the differences in treatment effects were usually slightly larger for the non-parametric methods than for the parametric and semi-parametric methods. (See Appendix Figure 1) Comparing the SEs for the Kaplan-Meier versus the Weibull model, the increases ranged from 0.3-24.4\%, and were lowest for RMS and highest for 2-year survival.

For piecewise constant hazards (scenario 2), biases were typically lowest for the non-parametric methods and when using a PWC time-varying Cox model. An exception to this was for RMS, where the na\"{i}ve Cox model and Weibull AFT model had the lowest bias. This likely occurred because bias from different timepoints were in different directions (positive at 2 and 5 years, negative at 10 years), and may have cancelled each other out when calculating the area under the survival curve. (Note that Figure 1 shows the absolute value of the average biases).  So, although biases may have been lower in this simulation for these mis-specified models, such behavior should not be relied upon in other scenarios.  Coverages (See Appendix Table 2) were close to the nominal when biases were low. Standard errors were generally similar to be base case, except for median survival, where they were somewhat larger.

When the effect of non-proportionality was modest (scenario 3), biases tended to be smaller than those of the base case, especially for estimates of survival probabilities at a given timepoint. When the treatment effect was modest (scenario 4), results were generally similar to those of the base case. In the small sample size case (scenario 5), the bias was similar to that of the base case, with the exception of RMS, where across methods, biases were larger than those of the base case. As expected, SEs were higher when the sample size was smaller. 

\begin{figure}
\caption{Mean bias (absolute value) in estimates from simulation studies}
\includegraphics[width=7in, height=5in]{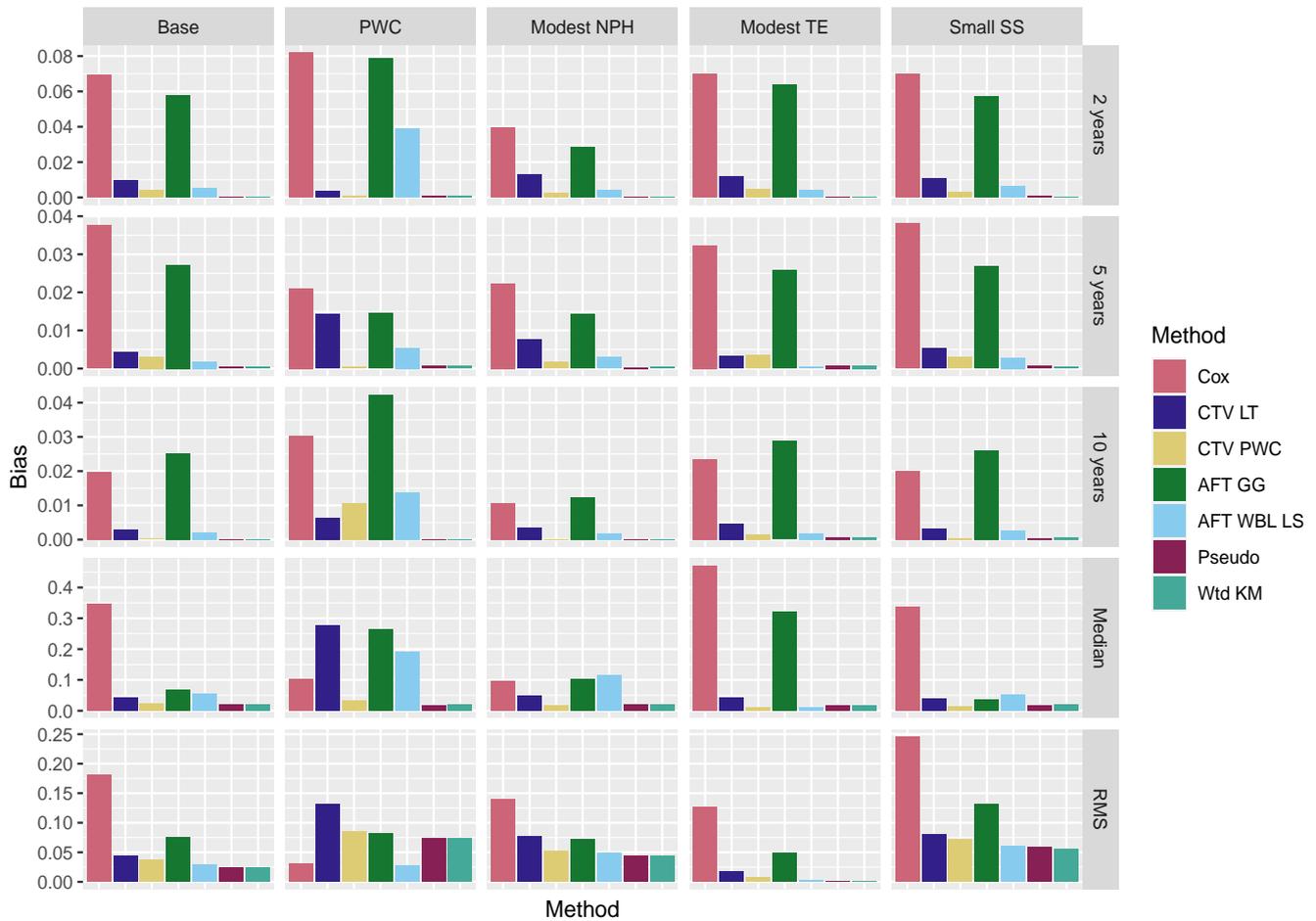}
\floatfoot{\footnotesize{CTV=Cox Time-Varying; LT=Log-Time; PWC=Piece-Wise Constant; AFT=Accelerated Failure Time; GG=Generalized Gamma; WBL LS= Weibull Location-Scale; Pseudo=Pseudo-Observations; Wtd KM=Weighted Kaplan-Meier; NHP=Non-Proportional Hazards; TE=Treatment effect; SS=Sample Size}}
\end{figure}

% Table generated by Excel2LaTeX from sheet 'Coverage_table'
\begin{table}[htbp]
  \centering
  \caption{Coverage for base case}
    \begin{tabular}{lrrrrr}
    \toprule
    \textbf{Method} & \multicolumn{1}{l}{\textbf{2y}} & \multicolumn{1}{l}{\textbf{5y}} & \multicolumn{1}{l}{\textbf{10y}} & \multicolumn{1}{l}{\textbf{Median}} & \multicolumn{1}{l}{\textbf{RMS}} \\
    \midrule
    Cox   & 0.00  & 0.07  & 0.61  & 0.81  & 0.71 \\
    CTV LT & 0.82  & 0.91  & 0.93  & 0.94  & 0.93 \\
    CTV PWC & 0.96  & 0.96  & 0.95  & 0.95  & 0.93 \\
    AFT GG & 0.00  & 0.37  & 0.44  & 0.94  & 0.90 \\
    AFT WBL LS & 0.89  & 0.93  & 0.93  & 0.95  & 0.93 \\
    Pseudo & 0.97  & 0.97  & 0.96  & 0.95  & 0.94 \\
    Wtd KM & 0.97  & 0.96  & 0.96  & 0.95  & 0.94 \\
    \bottomrule
    \end{tabular}%
  \label{tab:addlabel}%
\end{table}%

\subsection{Computational resources required}

The computational time and resources required to fit  each model varied substantially. Depending on the size of the dataset, computational considerations may become important, especially given our use of the bootstrap to compute confidence intervals.  The fastest method was weighted KM. To fit a Cox model with a time varying-effect of log-time, we found it  infeasibly slow to split the dataset at each observed failure time given our large sample size.  Instead, we limited the splits to occur at one-month intervals; we tested this approach and found that the results were nearly identical to those obtained by splitting at each failure time.  The pseudo observations approach was also relatively slow, and highly memory intensive, even when we estimated $S(t)$ at one-month intervals.  Table 2 shows the time required to fit each model once, and the total time required to calculate 500 bootstrap samples.   

\begin{table}[htbp]
  \centering
    \caption{Performance comparison: time required for N=5,000}
    \begin{tabular}{lrr}
    \toprule
    \textbf{Method} & \multicolumn{1}{l}{\textbf{Per bootstrap iteration (sec)}} & \multicolumn{1}{l}{\textbf{Total time (min)}} \\
    \midrule
    Cox & 0.98  & 8.21 \\
    CTV LT* & 7.60  & 63.44 \\
    CTV PWC & 0.41  & 3.40 \\
    AFT GG & 13.70 & 114.39 \\
    AFT Weibull LS & 5.26  & 43.90 \\
    Pseudo** & 2.17  & 18.14 \\
    Wtd KM & 0.14  & 1.18 \\
    \bottomrule
    \multicolumn{3}{l}{\footnotesize{*28 times higher if split at each failure}}        \\
    \multicolumn{3}{l}{\footnotesize{**60 times higher if estimated at each failure}}       \\
    \end{tabular}%
 \end{table}%

\section{Comparative effectiveness of Cancer Therapies}

We applied the previously described methods to estimate the effect of treatments on survival in two clinical studies using the National Cancer Database (NCDB). The NCDB is a large registry which encompasses approximately 70\% of newly diagnosed cancer cases in the United States.  It contains  variables describing patient and tumor characteristics, and the treatments they receive. \cite{boffa_using_2017}  Here, we describe two studies we conducted using the NCDB that featured two different forms of non-proportional hazards. The first is a moderately sized study where two treatment groups had similar survival initially, but differences in long-term outcomes. The second example is a much larger study where benefits of one treatment were attenuated over time.

\subsection{Sarcoma}

Soft-tissue sarcoma is a rare cancer of the connective tissue (e.g. fat, muscle, blood vessel), usually arising in the extremities. Stage III disease (high grade, large, or deep tumor) is typically treated surgically or with radiation therapy. It remains uncertain whether chemotherapy provides an overall survival benefit for these patients, and its use is optional according to national guidelines. \cite{von_mehren_margaret_nccn_2020}  We studied this question using the NCDB, creating a cohort of Stage III sarcoma patients (with various histologies) treated with definitive surgical resection of the primary tumor, with or without chemotherapy pre/post operatively.  We identified 5,337 cases with recorded overall survival data, of whom 28\% were treated with chemotherapy.  \cite{movva_patterns_2015}

Due to the non-randomized nature of the data, there were substantial differences between treatment groups.  For example, patients treated with chemotherapy were younger and had fewer recorded co-morbid conditions. Furthermore, the hazards for the treatment groups exhibited substantial non-proportionality, as is clearly evident upon examining the complementary log-log survival curves. (See Figure 2) Testing the Schoenfeld residuals provides further evidence of non-proportionality (P=0.047).  We used IPTW to account for covariate imbalance between the treatment groups.  The propensity score was estimated using all available covariates (including age, gender, insurance status, income, comorbidity score, tumor histology, grade, size, anatomic site, treating facility type, and travel distance.)  The IPTW Kaplan-Meier curves are shown in Figure 2.  In our original analysis, we also fit a time-varying Cox proportional hazards model, assuming a piecewise constant hazard ratio, which could change at two years.  We found that there was little effect of chemotherapy during the first two years after diagnosis, but that patients treated with chemotherapy did better in the long term. These results motivated simulation scenario 2 (PWC).

%Based on the shape of the adjusted curves, we then fit a time-varying Cox proportional hazards model, assuming assuming constant hazard ratios from zero to two years, and then from two years until the end of follow-up.  This analysis showed that there was little effect of chemotherapy during the first two years after diagnosis, but that patients treated with chemotherapy did better in the long term, perhaps indicating that treatment prevented disease recurrence.

%*** Include a brief "table 1" 

\begin{figure}
\caption{Survival in stage III sarcoma by chemotherapy}
\includegraphics[width=4in]{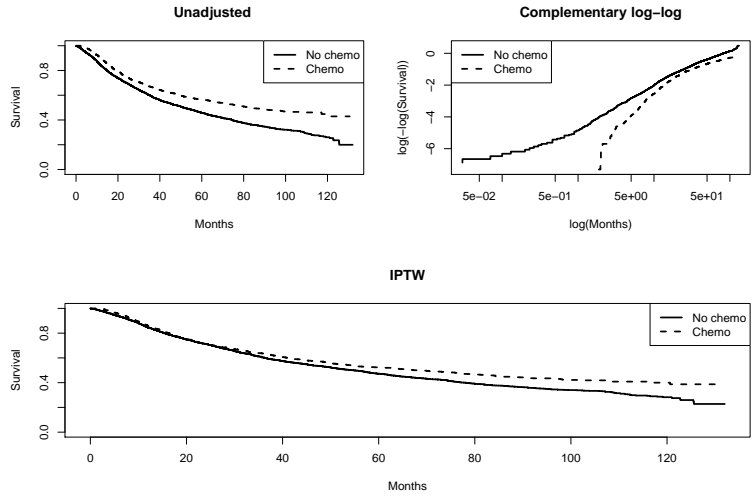}
\end{figure}

% Table generated by Excel2LaTeX from sheet 'GLM_final_res_04_10_20'
\begin{table}[htbp]
  \centering
  \caption{Chemotherapy in sarcoma: results from each model }
    \begin{tabular}{lrrrrrrrrr}
    \toprule
     & \multicolumn{3}{c}{\textbf{RMS}} & \multicolumn{3}{c}{\textbf{2 years}} &  \multicolumn{3}{c}{\textbf{5 years}} \\
         \cmidrule(lr){2-4}\cmidrule(lr){5-7} \cmidrule(lr){8-10}
    & \multicolumn{1}{c}{\textbf{$\Delta$} } & \textbf{LCL} & \textbf{UCL} & \multicolumn{1}{c}{\textbf{$\Delta$} } & \textbf{LCL} & \textbf{UCL} & \multicolumn{1}{c}{\textbf{$\Delta$} } & \textbf{LCL} & \textbf{UCL} \\
    \midrule
    Cox & \textbf{0.452} & 0.146 & 0.757 & \textbf{0.036} & 0.013 & 0.060 & \textbf{0.052} & 0.018 & 0.088 \\
    TV Cox: log-T & \textbf{0.457} & 0.132 & 0.752 & 0.018 & -0.011 & 0.043 & \textbf{ 0.052} & 0.019 & 0.087 \\
    TV Cox: PWC & \textbf{0.442} & 0.117 & 0.743 & -0.003 & -0.038 & 0.028 & \textbf{0.044} & 0.006 & 0.081 \\
    AFT Gen Gamma & \textbf{0.409} & 0.114 & 0.680 & \textbf{0.039} & 0.011 & 0.064 & \textbf{0.047} & 0.014 & 0.079 \\
    AFT Wbl TV shape & \textbf{0.468} & 0.160 & 0.773 & \textbf{0.026} & 0.002 & 0.048 & \textbf{0.054} & 0.020 & 0.090 \\
        Weighted K-M & \textbf{0.439} & 0.109 & 0.746 & -0.007 & -0.043 & 0.026 & \textbf{0.041} & 0.003 & 0.078 \\
    Pseudo-obs* & \textbf{0.439} & 0.110 & 0.745 & -0.007 & -0.043 & 0.026 & \textbf{0.041} & 0.003 & 0.078 \\
    \bottomrule
    \multicolumn{10}{l}{\footnotesize{*Centered weights}}\\
    \multicolumn{10}{l}{\footnotesize{$\Delta$ = Difference between ATEs for chemo vs no chemo, boldface denotes significance  }}\\
    \multicolumn{10}{l}{\footnotesize{LCL = Lower confidence limit, UCL = Upper confidence limit }}\\
    \end{tabular}%
  \label{tab:addlabel}%
\end{table}%

Here, we apply each of the methods discussed in the sections above to the sarcoma data. Results for RMS, 2 year survival, and 5 year survival are shown in Table 3.  Median and 10-year survival were not estimable from the observed data, so we excluded these outcomes.  We found that the different methods  yielded different effect size estimates; however, these differences were often modest and in some cases would not change statistical inferences. For example, $\Delta$RMS varied from a minimum of 0.409 (AFT with a generalized gamma distribution) to a maximum of 0.468, but all models showed an improvement in RMS for patients treated with chemotherapy.  In other cases the inferences would differ based on the model chosen.  At 2 years, the standard Cox model showed a difference of 3.6\% between the treated and control arms, which was statistically significant. The results of both AFT models also found significantly higher survival in the chemotherapy group. However, the weighted Kaplan-Meier method showed a non-significant difference of -0.7\%. Both time-varying Cox models and the pseudo-observations method also produced non-significant differences.     

%We did note one unexpected finding regarding the estimates found from pseudo-observations.  Using the standard IPTW weights produced estimates far different from the Kaplan-Meier estimated weights (e.g. 0.59 vs 0.44 for the RMS). ***Add CIs*** This was very surprising, as in simulations, the two method performed almost identically.  Further exploration of these results led us to conclude that the

We would like to note that in this data analysis example, we found that the pseudo-observations method was sensitive to deviations in the mean of the estimated weights from 1.  While the expectation of the mean of the weights is equal to 1, in practice there may be small deviations from the expected value.  The pseudo-observations method seemed to be more affected by these deviations than the other methods we evaluated. This sensitivity can be explained by how the weights are applied and how the survival function is estimated.  Each patient's (binary) contribution to the survival function is weighted, and the weighted indicators are then averaged.  Therefore, if the mean of the weights is not equal to 1, at baseline the survival function will not start at 1.  To address this issue in the estimation of the survival function when using pseudo-observations, we centered the weights so their mean was exactly 1.  However, we note that this was done heuristically, and not theoretically driven.  Further exploration of this issue is warranted.

\subsection{Renal Cancer}

The second motivating clinical example for this work was a study of surgical options for early stage renal cancer.  These small tumors can be treated with  Radical Nephrectomy (RN) or Partial Nephrectomy (PN). PN preserves more kidney tissue, but surgery/recovery is more difficult and it is more likely to be given to healthier patients. \cite{motzer_robert_j_nccn_2020, ristau_partial_2018} Here, we focus on a subgroup from the original study: patients aged 51-60 with T1a tumors, which gives a sample size of 28,973 patients (61.1\% treated with PN).  In unadjusted analysis, patients given PN had improved survival over those given RN.  However, these results are subject to confounding by indication.  Examination of the complementary log-log plots shows a notable deviation from non-proportionality, which is confirmed by testing the Schoenfeld residuals (P$<$0.0001). \cite{grambsch_proportional_1994}

When we fit an IPTW Cox PH model to these data, allowing the hazards to vary as a function of log-time, such that $h(t)  =  \lambda_0 (t) \mathrm{exp}(\beta_1 Z + \beta_2 Z \mathrm{log}(t))$  the maximum likelihood estimates are $\hat{\beta_1}=-0.468$ and $\hat{\beta_2}=0.136$.  Therefore, the non-proportionality observed here is close to our modest time-varying hazard simulation scenario (number 3), and the treatment effect is modest as well, similar to what we used in scenario 4. 
% Our simulation results showed that when time-varying hazards are modest, the bias in estimates from the standard (non-time-varying) Cox model was less pronounced than in the base case.  Nevertheless, it was larger than what occurred when time-varying Cox models or non-parametric (weighted Kaplan-Meier or Pseudo-observation) methods were used. These differences in biases were more pronounced for the two and five year survival estimates than for the RMS.  We also note that for the simulations with modest treatment effects, biases for the RMS tended to be smaller.

\begin{figure}
\caption{Survival in early stage renal cancer by surgery type}
\includegraphics[width=4in]{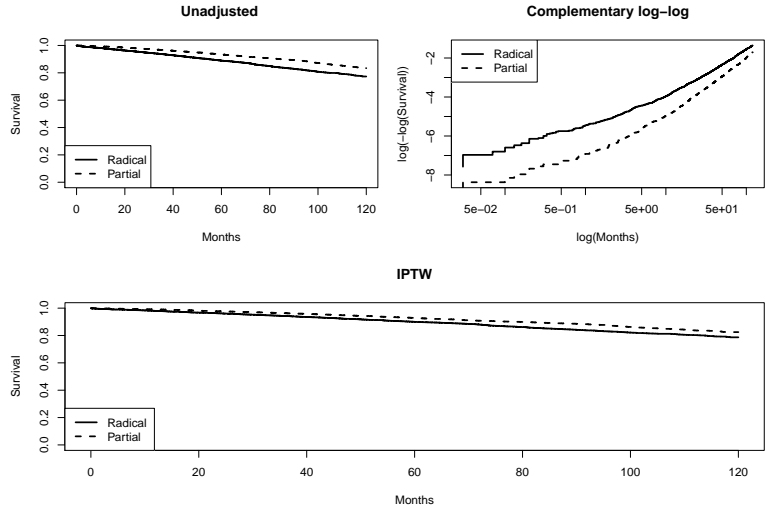}
\end{figure}

Our results were largely simiar, regardless of what model was used. (See Table 4) This was particularly the case for the RMS and five-year survival. Two-year survival showed the largest range in results, with estimated differences ranging from 0.9\% (Generalized Gamma AFT model) to 1.6\% (Cox model with piece-wise time-varying effects).  The standard Cox model gave an estimated difference of 1\%, compared to 1.5\% from the Kaplan-Meier method.  Even with these small effects, we found that none of the 95\% confidence limits crossed zero, so all models would lead to the same inference: a small benefit for PN over RN.  The congruous inferences were partially driven by the very large sample size, and the resulting small confidence intervals; however, given the clear evidence of non-proportionality, it is notable that the point estimates for the standard Cox model were close to those of time-varying or non-parametric methods.

% Table generated by Excel2LaTeX from sheet 'tmp'
\begin{table}[htbp]
  \centering
  \caption{Effect of PN vs RN in renal cancer: Results from each model}
    \begin{tabular}{lrrrrrrrrr}
    \toprule
       & \multicolumn{3}{c}{\textbf{RMS}} & \multicolumn{3}{c}{\textbf{2 years}} &  \multicolumn{3}{c}{\textbf{5 years}} \\
         \cmidrule(lr){2-4}\cmidrule(lr){5-7} \cmidrule(lr){8-10}
    & \multicolumn{1}{c}{\textbf{$\Delta$} } & \textbf{LCL} & \textbf{UCL} & \multicolumn{1}{c}{\textbf{$\Delta$} } & \textbf{LCL} & \textbf{UCL} & \multicolumn{1}{c}{\textbf{$\Delta$} } & \textbf{LCL} & \textbf{UCL} \\
          \midrule
    Cox   & \textbf{0.268} & 0.199 & 0.331 & \textbf{0.010} & 0.007 & 0.012 & \textbf{0.027} & 0.020 & 0.034 \\
    TV Cox: log-T & \textbf{0.267} & 0.199 & 0.330 & \textbf{0.014} & 0.010 & 0.018 &\textbf{ 0.029} & 0.022 & 0.036 \\
    TV Cox: PWC & \textbf{0.264} & 0.196 & 0.327 & \textbf{0.016} & 0.012 & 0.020 &\textbf{ 0.030} & 0.021 & 0.037 \\
    AFT Gen Gamma & \textbf{0.263} & 0.194 & 0.326 & \textbf{0.009} & 0.007 & 0.012 & \textbf{0.026} & 0.020 & 0.033 \\
    AFT Wbl TV shape & \textbf{0.264} & 0.196 & 0.329 & \textbf{0.019} & 0.014 & 0.023 & \textbf{0.031} & 0.024 & 0.038 \\
    Weighted K-M & \textbf{0.268} & 0.200 & 0.333 & \textbf{0.015} & 0.011 & 0.020 & \textbf{0.029} & 0.020 & 0.036 \\
    Pseudo-obs* & \textbf{0.269} & 0.200 & 0.334 & \textbf{0.015} & 0.011 & 0.020 & \textbf{0.029} & 0.020 & 0.036 \\
    \bottomrule
        \multicolumn{10}{l}{\footnotesize{*Centered weights}}\\
         \multicolumn{10}{l}{\footnotesize{$\Delta$ = Difference between ATEs for PN vs RN, boldface denotes significance  }}\\
    \multicolumn{10}{l}{\small{LCL = Lower confidence limit, UCL = Upper confidence limit }}\\
    \end{tabular}%
 % \label{tab:addlabel}%
\end{table}%

\section{Discussion}

When analyzing survival outcomes, we often see non-proportional hazards.  Based on our findings, it seems that non-proportionality is readily addressable in practice with weighted Kaplan-Meier curves, the simplest method we assessed. It performed quite well across a variety of scenarios and outcome measures of interest. The IPTW Kaplan-Meier method does not require specialized software or methods, and it also had the best computational performance. There is a penalty in terms of efficiency, but we found the increase in the size of the standard errors (compared to parametric methods) to be small in practice.  We believe that the  larger standard errors are a minor drawback when compared to the benefits of fewer assumptions and reduced bias. 

In this work, we used simple IPTW weights. Alternative propensity score weights have also been proposed; \cite{austin_moving_2015} notably, variance stabilized weights have good properties, especially when one treatment is given to a small proportion of the patients.  In our simulation studies, we used a simple functional form to model the covariates' relationship with the probability of receiving treatment. Therefore, the simple logistic regression model used to estimate the propensity scores was correctly specified.  In practice, there may be much more complex covariate effects, motivating more flexible procedures to estimate the propensity scores, such as ensemble machine learning methods. \cite{van_der_laan_mark_j_super_2007}  Such methods have been shown to improve prediction, but are often more resource-intensive to implement.

Our simulation studies were directly informed by the two cancer studies discussed in Section 5. We sought to better understand how choice of method may affect results of these real-world analyses, and to learn broader lessons about how best to accommodate non-proportionality in large, observational studies. Although both clinical examples had large sample sizes, we found in simulations that the results held up even when sample sizes were small. Taken together, our simulation results and clinical examples show that one can easily protect against incorrect inferences using IPTW Kaplan-Meier curves to estimate treatment effects.

\section*{Acknowledgement}
Research reported in this publication was supported by the National Cancer Institute of the National Institutes of Health under Award Number P30CA006927 and U54CA221705. The content is solely the responsibility of the authors and does not necessarily represent the official views of the National Institutes of Health. The data used  in  the  study  are  derived  from  a  de-identified  NCDB  file. The  American  College  of  Surgeons  and  the  Commission  on  Cancer  have  not  verified  and  are  not  responsible  for  the  analytic  or  statistical  methodology  employed,  or  the  conclusions  drawn  from  these  data by the investigator.  

\bibliographystyle{SageV} % "vancouver" reference style
\bibliography{Handorf_nonPH_IPTW_SMMR} % Entries are in the "refs.bib" file

\section*{Appendices}

\setcounter{table}{0}
\renewcommand{\thetable}{A\arabic{table}}

\begin{table}[htbp]
  \centering
  \caption{Simulation model coefficients ($\beta$)}
    \begin{tabular}{lrr}
        \toprule
          & \multicolumn{2}{c}{Coefficients} \\
          & \multicolumn{1}{c}{Treatment model} & \multicolumn{1}{l}{Survival model} \\
   \midrule
    Intercept & 0.529 & N/A \\
    age   & 0.03  & 0.04 \\
    charlson 1 & 0.15  & 0.3 \\
    charlson 2+ & 0.35  & 0.8 \\
    male  & 0.15  & 0.3 \\
    low.stage & -0.8  & -0.4 \\
    high.grade & 0.4   & 0.2 \\
    histology & 0.3   & 0.2 \\
    white & -0.4  & -0.1 \\
    hispanic & -0.1  & -0.2 \\
    facility 1 & -0.15 & 0 \\
    facility 2 & -0.2  & 0 \\
    facility 3 & -0.25 & 0 \\
    income 1 & -0.15 & -0.1 \\
    income 2 & -0.25 & -0.1 \\
    income 3 & -0.3  & -0.2 \\
    education 1 & -0.2  & -0.3 \\
    education 2 & -0.25 & -0.2 \\
    education 3 & -0.3  & -0.15 \\
    insurance 1 & -0.3  & -0.1 \\
    insurance 2 & -0.35 & -0.2 \\
        \bottomrule
    \end{tabular}%
 % \label{tab:addlabel}%
\end{table}%

\setcounter{figure}{0}
\renewcommand\thefigure{A\arabic{figure}}

\begin{figure}
\caption{Mean standard error of estimates from simulation studies}
\includegraphics[width=7in, height=5in]{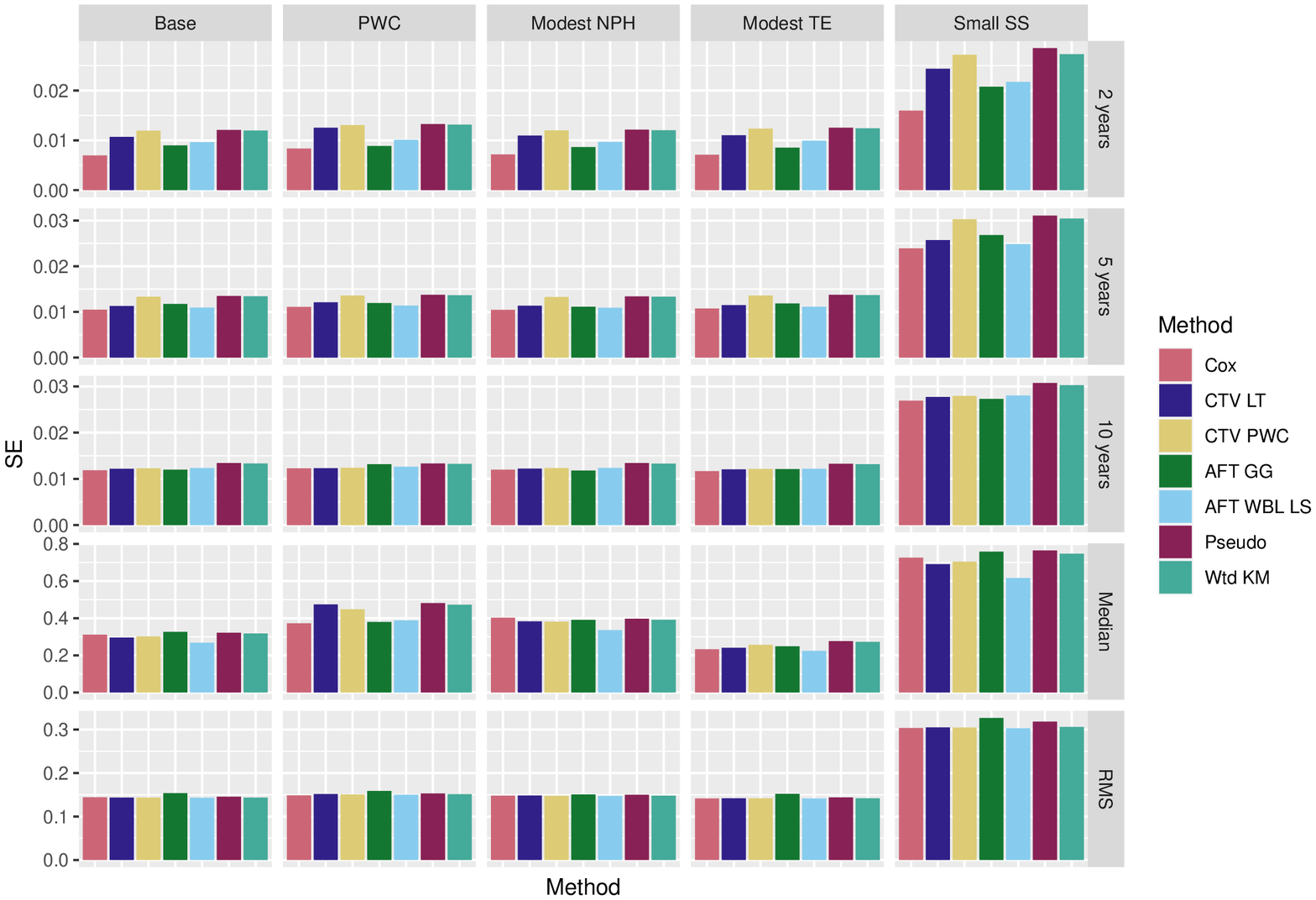}
\floatfoot{CTV=Cox Time-Varying; LT=Log-Time; PWC=Piece-Wise Constant; AFT=Accelerated Failure Time; GG=Generalized Gamma; WBL LS= Weibull Location-Scale; Pseudo=Pseudo-Observations; Wtd KM=Weighted Kaplan-Meier; NHP=Non-Proportional Hazards; TE=Treatment effect; SS=Sample Size}
\end{figure}

% Table generated by Excel2LaTeX from sheet 'Coverage_table'
\begin{table}[htbp]
  \centering
  \caption{Coverage probabilities by scenario}
    \begin{tabular}{llrrrrr}
    \toprule
    \textbf{Scenario} & \textbf{Method} & \multicolumn{1}{l}{\textbf{2y}} & \multicolumn{1}{l}{\textbf{5y}} & \multicolumn{1}{l}{\textbf{10y}} & \multicolumn{1}{l}{\textbf{Median}} & \multicolumn{1}{l}{\textbf{RMS}} \\
    \midrule
    PWC   & Cox   & 0.00  & 0.51  & 0.30  & 0.96  & 0.95 \\
    PWC   & CTV LT & 0.95  & 0.78  & 0.92  & 0.90  & 0.84 \\
    PWC   & CTV PWC & 0.96  & 0.96  & 0.87  & 0.94  & 0.90 \\
    PWC   & AFT GG & 0.00  & 0.75  & 0.10  & 0.90  & 0.91 \\
    PWC   & AFT WBL LS & 0.03  & 0.91  & 0.81  & 0.92  & 0.94 \\
    PWC   & Pseudo & 0.96  & 0.95  & 0.96  & 0.95  & 0.91 \\
    PWC   & Wtd KM & 0.96  & 0.95  & 0.96  & 0.95  & 0.92 \\
    Modest NPH & Cox   & 0.00  & 0.45  & 0.83  & 0.94  & 0.79 \\
    Modest NPH & CTV LT & 0.76  & 0.87  & 0.94  & 0.94  & 0.89 \\
    Modest NPH & CTV PWC & 0.96  & 0.97  & 0.96  & 0.95  & 0.92 \\
    Modest NPH & AFT GG & 0.13  & 0.72  & 0.80  & 0.95  & 0.89 \\
    Modest NPH & AFT WBL LS & 0.90  & 0.91  & 0.95  & 0.93  & 0.92 \\
    Modest NPH & Pseudo & 0.96  & 0.97  & 0.96  & 0.95  & 0.92 \\
    Modest NPH & Wtd KM & 0.97  & 0.98  & 0.97  & 0.95  & 0.92 \\
    Modest TE & Cox   & 0.00  & 0.15  & 0.48  & 0.49  & 0.86 \\
    Modest TE & CTV LT & 0.78  & 0.93  & 0.92  & 0.94  & 0.94 \\
    Modest TE & CTV PWC & 0.94  & 0.95  & 0.95  & 0.96  & 0.94 \\
    Modest TE & AFT GG & 0.00  & 0.42  & 0.35  & 0.75  & 0.94 \\
    Modest TE & AFT WBL LS & 0.91  & 0.95  & 0.94  & 0.95  & 0.95 \\
    Modest TE & Pseudo & 0.96  & 0.95  & 0.94  & 0.95  & 0.95 \\
    Modest TE & Wtd KM & 0.96  & 0.95  & 0.95  & 0.96  & 0.95 \\
    Small SS & Cox   & 0.02  & 0.64  & 0.88  & 0.93  & 0.86 \\
    Small SS & CTV LT & 0.93  & 0.93  & 0.95  & 0.97  & 0.93 \\
    Small SS & CTV PWC & 0.97  & 0.97  & 0.95  & 0.97  & 0.94 \\
    Small SS & AFT GG & 0.27  & 0.82  & 0.82  & 0.96  & 0.93 \\
    Small SS & AFT WBL LS & 0.91  & 0.93  & 0.95  & 0.96  & 0.95 \\
    Small SS & Pseudo & 0.96  & 0.98  & 0.96  & 0.96  & 0.96 \\
    Small SS & Wtd KM & 0.97  & 0.97  & 0.96  & 0.96  & 0.95 \\
    \bottomrule
    \end{tabular}%
  \label{tab:addlabel}%
\end{table}%

\end{document}